\documentclass[aps,prd,twocolumn,nofootinbib]{revtex4}

\usepackage{booktabs,makecell,multirow}

\usepackage{graphicx}
\usepackage{float}
\usepackage{color}

\usepackage[colorlinks=true, linkcolor=blue, citecolor=blue, urlcolor=blue]{hyperref}

\newcommand*{\RMN}[1]{\uppercase\expandafter{\romannumeral#1}}

\begin{document}

\title{The $P$-wave charmonium annihilation into two photons $\chi_{c0, c2}\rightarrow \gamma\gamma$ with high-order QCD corrections}

\author{Hua Zhou}
\email{zhouhua@cqu.edu.cn}

\author{Qing Yu}
\email{yuq@cqu.edu.cn}

\author{Xu-Dong Huang}
\email{hxud@cqu.eud.cn}

\author{Xu-Chang Zheng}
\email{zhengxc@cqu.edu.cn}

\author{Xing-Gang Wu}
\email{wuxg@cqu.edu.cn (corresponding author)}

\affiliation{Department of Physics, Chongqing University, Chongqing 401331, People's
Republic of China}

\date{\today}

\begin{abstract}

In this paper, we present a new analysis on the $P$-wave charmonium annihilation into two photons up to next-to-next-to-leading order (NNLO) QCD corrections by using the principle of maximum conformality (PMC). The conventional perturbative QCD prediction shows strong scale dependence and deviates largely from the BESIII measurements. After applying the PMC, we obtain a more precise scale-invariant pQCD prediction, which also agrees with the BESIII measurements within errors, i.e. $R={\Gamma_{\gamma\gamma}(\chi_{c2})} /{\Gamma_{\gamma\gamma}(\chi_{c0})}=0.246\pm0.013$, where the error is for $\Delta\alpha_s(M_\tau)=\pm0.016$. By further considering the color-octet contributions, even the central value can be in agreement with the data. This shows the importance of a correct scale-setting approach. We also give a prediction for the ratio involving $\chi_{b0, b2} \to\gamma\gamma$, which could be tested in future Belle II experiment.

\end{abstract}

\maketitle

Charmonium decays have been widely used to explore the interplay between the perturbative and non-perturbative dynamics due to its relatively clean platform, and they also play important roles in establishing the asymptotic freedom of quantum chromodynamics (QCD)~\cite{Appelquist:1974zd, DeRujula:1974rkb}. Among them, many attentions have been paid for the electromagnetic decays $\chi_{c0, c2}\to\gamma\gamma$. They have been measured by the CLEO and BESIII collaborations~\cite{Eisenstein:2001xe, Ablikim:2017rbk}; especially, in year 2017, the BESIII collaboration issued their measured value for the $R$-ratio,
\begin{eqnarray}
R_{\rm exp} &=&\frac{\Gamma_{\chi_{c2\to\gamma\gamma}}} {\Gamma_{\chi_{c0\to \gamma\gamma}}}=0.295\pm0.014\pm0.007\pm0.027,
\end{eqnarray}
where the errors are statistical, systematic, and the associated errors of the branching fraction ${\cal B}(\psi(3686)\to\gamma\chi_{c0, c2})$ and the total decay width $\Gamma_{\chi_{c0, c2}}$, respectively.

On the other hand, they have been calculated by using various approaches, such as the nonrelativistic potential model, the nonrelativistic QCD theory (NRQCD), the relativistic quark model, and the lattice QCD theory, respectively, c.f. Refs.\cite{Barbieri:1975am, Barbieri:1980yp, Li:1990sx, Gupta:1996ak, Munz:1996hb, Bodwin:1994jh, Huang:1996cs, Ambrogiani:2000vc, Ma:2002eva, Dudek:2006ut, Brambilla:2006ph, Sang:2015uxg} and references therein. Within the framework of NRQCD factorization theory, one has observed that the leading-order (LO) prediction is close to the experimental measurements, but this process is extremely sensitive to high-order QCD corrections and relativistic corrections, due to the fact that the typical magnitude of strong coupling constant and the squared relative velocity of the charm quark in charmonium, $\alpha_s(m_c)\sim v^2_c\sim 0.3$, are comparatively large. It is important to finish as more perturbative terms as possible so as to achieve a more accurate pQCD prediction. And in order to obtain a convincing fixed-order prediction, the influence of high-order correction on the $\chi_{c0, c2}\to\gamma\gamma$ must be carefully analyzed.

At the present, the QCD corrections to the $S$-wave heavy quarkonium electromagnetic/leptonic electromagnetic decays have been calculated up to next-to-next-to-leading-order (NNLO) level. The spin-singlet heavy quarkonium decays  $\eta_{c}\to\gamma\gamma$ and $\eta_{b}\to\gamma\gamma$ have been calculated up to NNLO level in Refs.\cite{Feng:2015uha, Czarnecki:2001zc}; and the spin-triplet heavy quarkonium decays $J/\psi\to e^+ e^-$ and $\Upsilon\to e^+ e^-$ decay have been calculated up to NNLO level by Refs.\cite{Czarnecki:1997vz, Beneke:1997jm}. In year 2016, the NNLO QCD corrections to the $P$-wave charmonium $\chi_{c0, c2}\to\gamma\gamma$ have been done by Ref.\cite{Sang:2015uxg}, which however show large renormalization scale dependence, and the predicted $R$-ratio cannot explain the above BESIII value. It is important to show what's the reason for such discrepancy.

Within the framework of NRQCD, one can factorize the decay width into non-perturbative matrix elements and perturbatively calculable short-distance coefficients, and the $R$-ratio becomes
\begin{equation}\label{Rratio}
R_c = \frac{(|A^{\chi_{c2}}_{1,1}|^2+|A^{\chi_{c2}}_{1,-1}|^2)}{5(|A^{\chi_{c0}}_{1,1}|^2)},
\end{equation}
where the helicity amplitude of this process is expressed by $A^{\chi_{cJ}}_{\lambda_1,\lambda_2}$, $\lambda=|\lambda_1-\lambda_2|$, $\lambda_{1,2}=\pm1$, $J=0,2$. The amplitude of the $P$-wave quarkonium electromagnetic decays $\chi_{c0, c2}\to\gamma\gamma$ can be expressed as~\cite{Sang:2015uxg}
\begin{equation}
A^{\chi_{c0, c2}}_{\lambda_1,\lambda_2} = C^{\chi_{c0, c2}}_{\lambda}(m_c,\mu_r,\mu_\Lambda) \frac{\left\langle0\vert\chi^{^\dag}K_{3_{P_{0,2}}}\Psi \vert \chi_{c_{0,2}} \right\rangle_{\mu_\Lambda}}{m^{3/2}_c},
\label{eqf}
\end{equation}
where the color-singlet $P$-wave long-distance matrix element can be related to the first derivative of the radial wave function at the origin,
\begin{equation}
\left\langle0\vert\chi^{^\dag} K_{3_{P_{0,2}}}\Psi \vert\chi_{c_{0,2}}\right\rangle_{\mu_\Lambda} =\sqrt\frac{3 N_c}{2\pi} \; \left.\overline{R}^\prime_{\chi_{c0, c2}}(0)\right|_{\mu_\Lambda},
\label{eqw}
\end{equation}
where $N_{c}=3$ is the ${\rm SU}_{c}(3)$ color number, $\overline{R}^{\prime}_{\chi_{c0, c2}}(0)$ are first derivatives of $\chi_{c0, c2}$ radial wavefunctions at the origin. The spin-splitting effect for the radiation wavefunctions of $\chi_{c0, c2}$ are small, and by defining the $R$-ratio (\ref{Rratio}), the uncertainties caused by the matrix elements can be greatly suppressed. The perturbative part $C^{\chi_{c0, c2}}_{\lambda}(m_c, \mu_r, \mu_\Lambda)$ up to NNLO level can be read from Ref.\cite{Sang:2015uxg}, where $\mu_r$ and $\mu_\Lambda$ are renormalization and factorization scales, respectively.

In dealing with the perturbative series of $R$-ratio, one usually sets $\mu_r = m_c$ so as to eliminate large logarithmic terms in powers of $\ln\mu_r^2/m_c^2$, and then varies it within certain range to ascertain the uncertainty. This simple method causes the mismatching of $\alpha_s$ with its perturbative coefficients at each order, breaks the renormalization group invariance~\cite{Wu:2014iba}, and leads to conventional renormalization scheme-and-scale ambiguities. Such ambiguities could be softened to a certain degree by including higher-order terms. However due to its complexity, the exact NNNLO corrections of $\chi_{c0, c2}\to\gamma\gamma$ shall not be available in near future, thus it is important to find a correct way to achieve a reliable and accurate prediction by using the known NNLO series.

The renormalization scale-setting problem is one of the most important issues for pQCD theory, which has a long history, cf. the review~\cite{Wu:2013ei}. To solve it, we adopt the single-scale approach~\cite{Shen:2017pdu} of the principle of maximum conformality (PMC) to analyze the decay width of $\chi_{c0,2}\to\gamma\gamma$ up to NNLO QCD corrections. By using the renormalization group equation recursively, the PMC determines the precise $\alpha_s$ value of the process by using the non-conformal $\beta$-terms in pQCD series~\cite{Brodsky:2011ta, Brodsky:2011ig, Brodsky:2012rj, Brodsky:2013vpa, Mojaza:2012mf}. After applying the PMC, the resultant pQCD series becomes conformal, the magnitude of $\alpha_s$ and the perturbative coefficients become well matched, and then we obtain exact values for each order. The PMC predictions are renormalization scheme-and-scale independent~\cite{Wu:2018cmb}, thus the conventional scale-setting ambiguities is eliminated. Due to the perturbative nature of the pQCD theory, there is residual scale dependence because of unknown higher-order terms~\cite{Zheng:2013uja}; For the PMC perturbative series, such residual scale dependence shall generally be highly suppressed even for lower-order predictions~\cite{Wu:2019mky}.

One can rewrite the $R$-ratio (\ref{Rratio}) of $\chi_{c0,2}\to\gamma\gamma$ as the following perturbative form,
\begin{eqnarray}
R_c &=&\Omega[1+r_{1}a_s(\mu_r)+r_{2}a^2_s(\mu_r)+{\cal O}(a_s^3)],
\label{eqs}
\end{eqnarray}
where $a_s={\alpha_s}/{4\pi}$ and $\Omega=|{\overline{R}^{\prime}_{\chi_{c2}}} / {\overline{R}^{\prime}_{\chi_{c0}}}|^2 \simeq 1$. The perturbative coefficients $r_i$ can be derived into conformal terms $r_{i,0}$ and non-conformal terms $r_{i,j\neq0}$ by using the degeneracy relations among different orders~\cite{Bi:2015wea}, i.e.,
\begin{eqnarray}
r_1&=&r_{1,0} ,\\
r_2&=&r_{2,0}+r_{2,1}\beta_0 ,
\end{eqnarray}
where $\beta_0=11-\frac{2}{3}n_f$, representing the one-loop $\beta$-function, in which $n_f$ is the active flavor numbers. Using the NNLO results given in Ref.\cite{Sang:2015uxg}, we obtain
\begin{eqnarray}
r_{1,0}&=&-4.40967C_F,\\
r_{2,0}&=&47.0754C_{F}^2-\frac{9.23656C_F T_F}{e^{2}_{c}}\nonumber\\
&&-(22.7351)C_F n_H T_F\nonumber\\
&&+C_A C_F(-34.8488+123.208T_F)\nonumber\\
&&+C_F(-2.84012\times10^{-15}C_A\nonumber\\
&&+37.8993C_F)\ln\frac{\mu_\Lambda}{m_c},\\
r_{2,1}&=&-33.6021C_F T_F-4.40967C_F\ln[\frac{\mu^2_r}{m^2_c}],
\end{eqnarray}
where $n_H=1$, $C_{F}=\frac{N^{2}_{c}-1}{2N_{c}}$, $C_{A}=N_{c}$, $T_F=1/2$, and $e_{c}$ represents the charm-quark electric charge.

After applying the standard procedures of the PMC single-scale approach~\cite{Shen:2017pdu} to the pQCD series (\ref{eqs}), we obtain the following conformal series,
\begin{eqnarray}
R_c &=&\Omega[1+r_{1,0}\alpha_{s}(Q_\ast)+r_{2,0}\alpha^{2}_{s}(Q_\ast)],
\end{eqnarray}
where $Q_\ast$ is the PMC scale, which is obtained by requiring all non-conformal items vanish. It is the effective scale which replaces the individual PMC scales at each order in PMC multi-scale approach~\cite{Brodsky:2013vpa, Mojaza:2012mf} in the sense of a mean value theorem. At present, by using the known NNLO perturbation series, the PMC scale can be fixed at the leading-log accuracy:
\begin{eqnarray}
\ln\frac{Q^2_\ast}{m^2_c}&=&-\frac{\hat r_{2,1}}{\hat r_{1,0}}+{\cal O}(\alpha_s),
\label{equ1}
\end{eqnarray}
where $\hat r_{i,j}=r_{i,j}|_{\mu_{r}=m_c}$. It is noted that $Q_\ast$ is independent to any choice of renormalization scale $\mu_r$, together with the scale-invariant conformal coefficients, the conventional renormalization scale ambiguity is eliminated. Using Eq.(\ref{equ1}), we obtain $Q_\ast=0.250$~GeV, which is close to the QCD asymptotic scale.

Because the effective momentum flow $Q_\ast$ of the process is close to the QCD asymptotic scale $\Lambda$, we need to choose a low-energy model for $\alpha_s$ so as to achieve a reliable prediction. In the literature, a variety of low-energy $\alpha_s$ models have been suggested~\cite{Shirkov:2012ux, Webber:1998um, Cornwall:1981zr, Godfrey:1985xj, Halzen:1992vd, Badalian:2001by, Brodsky:2010ur, Shirkov:1997wi, Shirkov:2004ar}. A comparison of various low-energy $\alpha_s$ models has been given in Ref.\cite{Zhang:2014qqa}. In the present paper, for clarity, we adopt the CON model to do our discussion. It is derived from continuum theory~\cite{Halzen:1992vd} and uses the exchanged gluons with an effective dynamical mass $m_g$, and determines the non-perturbative dynamics of gluons by using the Dyson-Schwinger equation. More explicitly, the CON low-energy $\alpha_s$ model is expressed as follows:
\begin{eqnarray}
\alpha^{\rm CON}_{s}(\mu)&=&\frac{\pi}{\beta_{0}\ln{\frac{4M^{2}_{g}+\mu^{2}}{\Lambda^{2}}}},
\end{eqnarray}
where $M^{2}_{g}=m^{2}_{g}\left(\frac{\ln{(\mu^{2}/\Lambda^{2} +4m^{2}_{g}/\Lambda^{2})}}{\ln(4m^{2}_{g}/\Lambda^{2})}\right)^{-12/11}$, and $m_{g}=500\pm200~\rm MeV$~\cite{Cornwall:1981zr}. The asymptotic scale $\Lambda$ can be fixed by using the $\alpha_s$ measured at a typical energy scale. More definitely, by using $\alpha^{\overline{\rm MS}}_{s}(M_{\tau})=0.325\pm0.016$~\cite{Zyla:2020zbs}, which leads to $\Lambda|_{n_{f}=3}=0.383^{+0.029}_{-0.031}$ GeV, $\Lambda|_{n_{f}=4}=0.324^{+0.029}_{-0.029}$ GeV, and $\Lambda|_{n_{f}=5}=0.223^{+0.022}_{-0.023}$ GeV.

\begin{figure}[htb]
\centering
\includegraphics[width=0.48\textwidth]{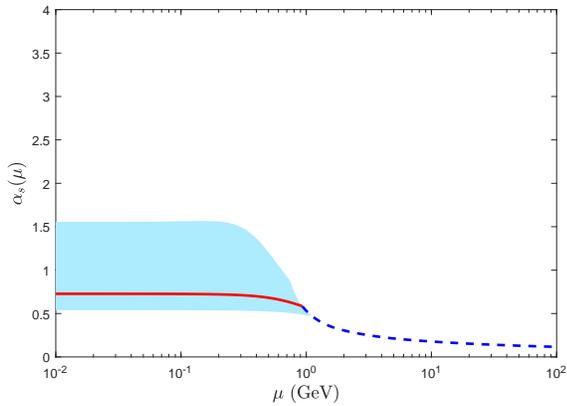}
\caption{The strong coupling constant $\alpha_s$ within different energy scales. In low-energy region, the shaded band is for CON-model with $m_{g}=500^{+200}_{-200}$~MeV. In large-energy region, the conventional $\overline{\rm MS}$ $\alpha_s$ running behavior is adopted. $\Lambda|_{n_{f}=3}=0.383$ GeV. }
\label{figo}
\end{figure}

Fig.\ref{figo} shows the $\alpha_s$-running behavior at different scales, the $\alpha_s$ CON-model with $m_{g}=500^{+200}_{-200}$~MeV is adopted in low-energy region which is shown by shaded band. The smooth connection between the low-energy region and the large-energy region is obtained by using the matching scheme proposed in Ref.\cite{Deur:2014qfa}. More explicitly, by requiring the first derivatives of $\alpha_s$ to be the same at the crossing point of the two energy regions, the $\alpha_s$ transition scale is determined to be $0.933^{+0.183}_{-0.191}$ GeV.

\begin{table}[htb]
\begin{center}
\begin{tabular}{c c c  c  c c c c c}
\hline
& ~$R_{c}$~ & ~LO-terms~ & ~NLO-terms~ & ~NNLO-terms~ & ~Total~\\
\hline
& $\rm Conv.$ & 0.267 & -0.157$^{-0.092}_{+0.043}$ & -0.043$^{+0.026}_{-0.005}$ & 0.067$^{-0.066}_{+0.038}$\\
& $\rm PMC$ & 0.267 & -0.250 & 0.229 & 0.246\\
\hline
\end{tabular}
\caption{Contributions from each loop terms for $R_c$ up to NNLO level under conventional (Conv.) and PMC scale-setting approaches, respectively. The PMC predictions are scale invariant, and the Conv. predictions are highly scale dependent, whose central values are for $\mu_r=m_{c}$ and the errors are for $\mu_r\in[1\;{\rm GeV}, 2m_c]$. $\mu_{\Lambda}=1$ GeV.  }
\label{tabw}
\end{center}
\end{table}

To do the numerical calculation, we take the $c$-quark pole mass $m_c=1.68$ GeV~\cite{Tanabashi:2018oca}, and set the factorization $\mu_{\Lambda}=1$ GeV. In Table \ref{tabw}, we give the contributions of each loop terms of $R_c$ under conventional and PMC scale-setting approaches, respectively. The conventional predictions are highly $\mu_r$-dependent for each loop terms and the total contributions of $R_c$, e.g. as shown by the following Eqs.(\ref{equ112}-\ref{equ113}), the renormalization scale uncertainty for $R^{\rm Conv.}_{\rm c, total}$ within the range of $\mu_r\in[1{\rm GeV}, 2m_c]$ is about $(^{-99\%}_{+57\%})$:
\begin{eqnarray}
\label{equ112}
R^{\rm Conv.}_{\rm c, total}|_{\mu_{r}=1\;{\rm GeV}} &=& 0.001~{\rm GeV},\\
R^{\rm Conv.}_{\rm c, total}|_{\mu_{r}=m_c} &=& 0.067~{\rm GeV},\\
R^{\rm Conv.}_{\rm c, total}|_{\mu_{r}=2m_c}&=& 0.105~{\rm GeV}.
\label{equ113}
\end{eqnarray}
Those values deviate from the BESIII measurement~\cite{Ablikim:2017rbk} by at least $\sim 3.9\sigma$. Moreover, the separate scale uncertainties for NLO-terms and NNLO-terms are ($^{+59\%}_{-27\%}$) and ($^{-60\%}_{+12\%}$), respectively. After applying the PMC, the conventional scale uncertainty is removed, i.e. we obtain $R^{\rm PMC}_{\rm c, total}\equiv 0.246$ for any choice of $\mu_r$, which agrees with the BESIII measurement within errors. By further taking the $\alpha_s$ shift, $\Delta\alpha_s(M_\tau)=\pm0.016$, into consideration, we obtain $\Delta R^{\rm Conv.}_{\rm c, total}|_{\mu_{r}=1\;{\rm GeV}} =\left(^{+0.026}_{-0.029}\right)$, $\Delta R^{\rm Conv.}_{\rm c, total}|_{\mu_{r}=m_c}=\pm 0.013$, $\Delta R^{\rm Conv.}_{\rm c, total}|_{\mu_{r}={\rm 2m_c}}=\pm 0.007$, and $\Delta {R^{\rm PMC}_{\rm c, total}}=\pm0.013$.

\begin{figure}[htb]
\centering
\includegraphics[width=0.48\textwidth]{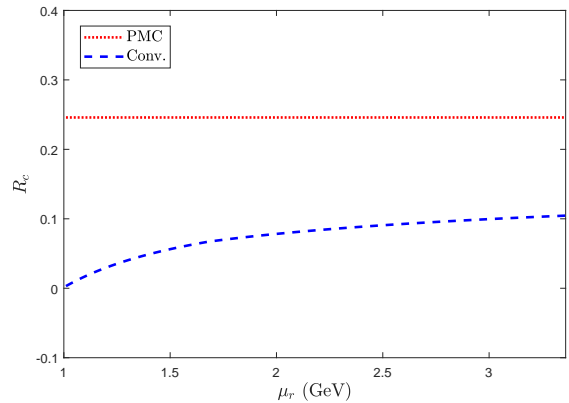}
\caption{The NNLO $R_c$-ratios under conventional and PMC scale-setting approaches as a function of $\mu_r$. $\mu_\Lambda=1$ GeV. }
\label{figw}
\end{figure}

Fig.~\ref{figw} shows explicitly how the $R_{c}$-ratio changes with different choices of $\mu_r$. It shows that after applying the PMC, the perturbative series is independent to any choice of $\mu_r$, and the conventional renormalization scale ambiguity is removed. The PMC conformal series ensures the scheme independence of the pQCD prediction, and together with the scale invariance, its behavior indicates the intrinsic perturbative behavior of the series. The NNLO conformal coefficient $r_{2,0}=126.74$ is larger than the conventional coefficient $r_2=-59.93^{+50.84}_{-67.92}$ for $\mu_{r}\in[1{\rm GeV}, 2m_c]$, this explains why the PMC magnitude of the NNLO-terms is larger than the conventional one. It is noted that the smaller conventional NNLO coefficient $r_{2}$ is due to accidental cancelation of conformal and non-conformal terms, which is however highly scale dependent, leading to scale dependent series.

\begin{figure}[htb]
\centering
\includegraphics[width=0.48\textwidth]{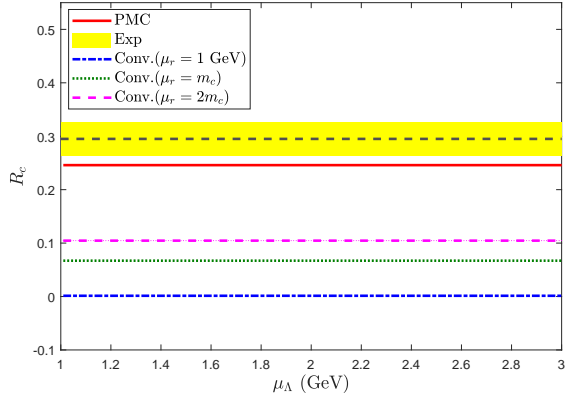}
\caption{The NNLO $R_c$-ratio versus the factorization scale $\mu_\Lambda$ under conventional and PMC scale-setting approaches. Three typical $\mu_r$ are adopted, and the PMC prediction is independent to those choices of $\mu_r$. The shaded band represents the BESIII data~\cite{Ablikim:2017rbk} with the dashed line being its central value. }
\label{figoD}
\end{figure}

A physical observable should not dependent on the choices of manly introduced parameters such as the renormalization scale and the factorization scale. In the above, we have shown that by applying the PMC, the NNLO $R$-ratio removes the conventional renormalization dependence. Due to the heavy quark spin symmetry, the matrix elements for $\chi_{c0}$ and $\chi_{c2}$ are the same, and $R$-ratio avoids the uncertainties caused by different choices of matrix elements. However, because the NNLO-coefficient $r_{2,0}$ is explicitly $\mu_\Lambda$-dependent, the $R$-ratio shows explicit $\mu_\Lambda$ dependence, whose magnitude is large~\cite{Sang:2015uxg}. We observe that such large factorization scale dependence could be removed by taking the evolution effects of the matrix element into consideration; That is, because the anomalous dimensions for $\chi_{c0}$ and $\chi_{c2}$ are different~\cite{Hoang:2006ty}, those two matrix elements are different at different $\mu_\Lambda$, which could compensate the $\mu_\Lambda$-dependence from the hard part. Using the evolution equation~\cite{Bodwin:1994jh}, we obtain
\begin{eqnarray}
\ln {\frac{\langle{\cal{O}}_{1}(^{3}P_{0})\rangle|_{\mu_{\Lambda}}}{\langle{\cal{O}}_{1}(^{3}P_{0})\rangle |_{\mu_{\Lambda 0}}}}&=&\frac{C_{F}(4C_{F}+C_{A})}{6} \alpha^{2}_{s}\ln{\frac{\mu^{2}_{\Lambda}}{\mu^{2}_{\Lambda0}}},\\
\ln {\frac{\langle{\cal{O}}_{1}(^{3}P_{2})\rangle|_{\mu_{\Lambda}}}{\langle{\cal{O}}_{1}(^{3}P_{2})\rangle |_{\mu_{\Lambda 0}}}}&=&\frac{C_{F}(13C_{F}+C_{A})}{60} \alpha^{2}_{s}\ln{\frac{\mu^{2}_{\Lambda}}{\mu^{2}_{\Lambda0}}}.
\end{eqnarray}
As required, the difference is an ${\cal O}(\alpha_s^2)$-order effect. If taking ${\cal{O}}_{1}(^{3}P_{J})|_{\mu_{\Lambda0}=1\rm GeV}=0.107~\rm GeV^{5}$~\cite{Braaten:2002fi} as the initial value, we obtain ${\cal{O}}_{1}(^{3}P_{0})|_{\mu_{\Lambda}=\rm 3 GeV}=0.139~\rm GeV^{5}$ and ${\cal{O}}_{1}(^{3}P_{2})|_{\mu_{\Lambda}=\rm 3 GeV}=0.124~\rm GeV^{5}$. Though the differences are small, we observe that the net factorization scale dependence of $R_c$ can be removed. This can be explicitly shown by Fig.\ref{figoD}, in which the flat lines for $R_c$ versus $\mu_\Lambda$ indicates that the $R_c$-ratio is independent to any choice of $\mu_\Lambda$. Fig.\ref{figoD} shows that after applying the PMC, the predicted $R_c$ becomes more closer to the BESIII value, which still has a slight gap from the data.

\begin{figure}[htb]
\centering
\includegraphics[width=0.48\textwidth]{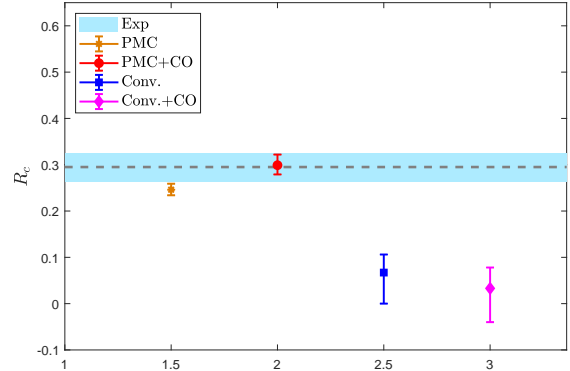}
\caption{The NNLO $R_c$-ratios with or without color-octet (CO) contributions under conventional and PMC scale-setting approaches. The error bars are for $\Delta\alpha_{s}(M_\tau)=\pm0.016$ and $\mu_r\in[1{\rm GeV}, 2m_c]$. The dashed line is the central value of BESIII and the shaded band represents its error~\cite{Ablikim:2017rbk}. }
\label{figo11}
\end{figure}

According to Refs.\cite{Ma:2002eva, Brambilla:2017kgw, Xu:2014zra}, contributions from the color-octet (CO) components in charmonium should not be ignored. For example, it has been argued that the color-octet components shall shift the decay width by $\Delta\Gamma^{\rm CO}_{\chi_{c0}}\simeq -0.3$ GeV and $\Delta\Gamma^{\rm  CO}_{\chi_{c2}} \simeq -0.227$ GeV~\cite{Ma:2002eva}. If taking those extra color-octet contributions into consideration, we obtain $R^{\rm PMC}_{\rm c, total}=0.246$ to $0.299$. Fig.~\ref{figo11} shows the $R_c$-ratio with or without the color-octet contributions, where the error bars are for $\Delta\alpha_{s}(M_\tau)=\pm0.016$ and $\mu_r\in[1{\rm GeV}, 2m_c]$. These results show a better match with the experimental data. Thus the CO contributions should be taken into consideration for a sound prediction.

\begin{table}[htb]
\begin{center}
\begin{tabular}{  c c c  c  c c c c c }
\hline
& ~$ R_{b}$~          & ~$\rm{LO}$~        & ~$\rm{NLO}$~    & ~$\rm{NNLO}$~             & ~$\rm{Total}$~\\
\hline
& $\rm Conv.$  & 0.267   &   -0.101$^{-0.148}_{+0.017}$  & -0.026$^{+0.120}_{-0.005}$    & 0.140$^{-0.028}_{+0.012}$\\
& $\rm PMC$  &   0.267    &  -0.237     &  0.135      & 0.165\\
\hline
\end{tabular}
\caption{Contributions from each loop terms for $R_b$ up to NNLO level under conventional (Conv.) and PMC scale-setting approaches, respectively. The PMC predictions are scale invariant, and the Conv. predictions are highly scale dependent, whose center values are for $\mu_r=m_{b}$, $\mu_{\Lambda}=m_{b}$.  and the errors are for $\mu_r\in[1\;{\rm GeV}, 2m_b]$.}
\label{tabfff}
\end{center}
\end{table}

\begin{figure}[H]
\centering
\includegraphics[width=0.48\textwidth]{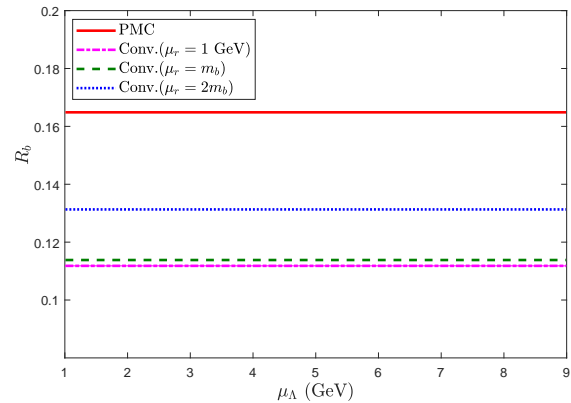}
\caption{The NNLO $R_b$-ratios versus the factorization scale $\mu_\Lambda$ under conventional and PMC scale-setting approaches. Three typical $\mu_r$ are adopted, and the PMC prediction is independent to those choices of $\mu_r$. }
\label{figo111}
\end{figure}

As a final remark, the above analysis can be conveniently applied for the $P$-wave bottomonium decays to two photons, which could be measured by the future high precision Belle-II experiment. We present the NNLO $R_b$-ratio under conventional and PMC scale-setting approaches in Table~\ref{tabfff}. To do the numerical calculation, we take the $b$-quark pole mass $m_b=4.78$ GeV~\cite{Tanabashi:2018oca}. Due to $\alpha_s(m_b)\sim 0.1$, a better pQCD convergence is observed for the bottomonium case, and the scale uncertainty is smaller, i.e. the net NNLO scale error is about $29\%$ for $\mu_r\in[1\;{\rm GeV}, 2m_b]$.

As a summary, in the present paper, we have studied the $R_c$-ratio up to NNLO accuracy. Under conventional scale-setting approach, the renormalization scale uncertainty is large, which is about $(^{-99\%}_{+57\%})$ for $\mu_r\in[1\;{\rm GeV}, 2m_c]$. By applying the PMC, we obtain a more accurate pQCD prediction without renormalization scale uncertainty, $R_c|_{\rm PMC}=0.246\pm0.013$, whose error is caused by $\Delta\alpha_s(M_\tau)=\pm 0.016$. This prediction agrees with the latest BESIII data within errors.

\hspace{2cm}

\noindent {\bf Acknowledgments:} This work was supported in part by the Natural Science Foundation of China under Grant No.11625520, No.12047564 and No.12005028, by the Fundamental Research Funds for the Central Universities under Grant No.2020CQJQY-Z003, and by the Chongqing Graduate Research and Innovation Foundation under Grant No.ydstd1912.

\end{document}